\documentclass[aps,preprint,preprintnumbers,amsmath,amssymb,nofootinbib]{revtex4}
\usepackage{amssymb}
\usepackage{lscape}
\usepackage{amsmath}
\usepackage{bm}
\usepackage{graphicx}
\usepackage{epsfig}
\begin{document}

\title{Inflationary expansion of the universe with variable timescale}
\author{$^{2}$ Juan Ignacio Musmarra\footnote{jmusmarra@mdp.edu.ar}, $^{1,2}$ Mariano Anabitarte\footnote{anabitar@mdp.edu.ar},  $^{1,2}$ Mauricio Bellini
\footnote{{\bf Corresponding author}: mbellini@mdp.edu.ar} }
\address{$^1$ Departamento de F\'isica, Facultad de Ciencias Exactas y
Naturales, Universidad Nacional de Mar del Plata, Funes 3350, C.P.
7600, Mar del Plata, Argentina.\\
$^2$ Instituto de Investigaciones F\'{\i}sicas de Mar del Plata (IFIMAR), \\
Consejo Nacional de Investigaciones Cient\'ificas y T\'ecnicas
(CONICET), Mar del Plata, Argentina.}

\begin{abstract}
We explore a cosmological model in which the time scale is variable with the expansion of the universe and the effective spacetime is driven by the inflaton field. An example is considered and their predictions are contrasted between Planck 2018 data. We calculate the spectrum indices and the slow-rolling parameters of the effective potential. The results are in very good agreement with observations.
\end{abstract}
\maketitle

\section{Introduction and motivation}

The emergence of quantum spacetime in the universe has been subject of study in the last years\cite{var}. This is an intriguing issue in the history of the universe that remains unsolved\cite{xz}\cite{b}. Possible dissipative effects in the context of fundamental theories of gravity have been discussed\cite{lc}, as well as extra-dimensional models where a condensate of fermion fields drives the expansion of the universe\cite{..}. Other proposals have been studied in a causal extra-dimensional set theory\cite{da}. On the other hand, the Friedmann-Robertson-Walker (FRW) metric is used to describe the large-scale expansion of the universe, and it can be considered as an effective (or phenomenological) way to describe a more fundamental line element which has a quantum mechanical origin\cite{mb}. In this metric the time scale along all the expansion of the universe is considered as a constant. However, it is sensate to think that this scale of time was not always the same. If the gravitational field was very intense at the beginning of the universe, it is reasonable to think that events spatially very close, could be subjected to a more slow rate of temporal flux. This means that with the expansion of the universe, spatially very separated events should be described by a more intense flux of time, as the gravitational field becomes weaker and weaker with the expansion of the universe.

In this work we shall consider, in a phenomenological approach, that the emergent spacetime grows exclusively due to the energy density transferred by the inflaton field at cosmological scales. For generality, we shall use a metric in which time scales of events at cosmological scales are not the same during the expansion of the emerging primordial universe. This means that scaling of time will be considered as variable along the expansion.

\section{The Model}

We consider an expanding universe that is spatially flat, isotropic and homogenous, which is not necessary in a vacuum background. The background metric is described by the line element
which describes a non-vacuum Friedman-Lemaitre-Robertson-Walker (FLRW) metric
\begin{equation}\label{back}
d{S}^2 = {g}_{\mu\nu} d{x}^{\mu} d{x}^{\nu}= e^{-2\int \Gamma(t)\, \,dt}dt^2 - a_0^2 \,\, e^{2\int H(t) dt}\,\, {\delta}_{ij}\, \,d{x}^i d{x}^j,
\end{equation}
such that $H(t)$ is the Hubble parameter on the background metric and $\Gamma(t)$ describes the time scale of the background metric. This should be the case in an emergent accelerated universe in which the spacetime is growing and the time scale can be considered variable with the expansion. In this paper we shall consider natural units, so that $c=\hbar=1$. The case of a background expansion on a vacuum is recovered by setting $\Gamma=0$.

To describe the expansion of the universe, we consider a single scalar field $\phi$ which is minimally coupled to gravity and drives the expansion
\begin{equation}\label{ref}
{\cal I} = \int d^4x \, \sqrt{-\hat{g}} \,\left\{ \frac{{\cal R}}{16\pi G}+\tilde{{\cal L}}_{\phi}\right\},
\end{equation}
where $\hat{g}$ is the determinant of $g_{\alpha\beta}$ and $\tilde{{\cal L}}_{\phi}$ the Lagrangian density for the scalar field:
\begin{displaymath}
\tilde{{\cal L}}_{\phi}= -\left[\frac{1}{2} g^{\alpha\beta} \phi_{,\alpha}\phi_{,\beta} - V(\phi)\right].
\end{displaymath}
If we use the definition for the stress tensor: $\tilde{T}_{\mu\nu}= 2 \frac{\delta\tilde{{\cal L}}_{\phi}}{\delta g^{\mu\nu}}- g_{\mu\nu}\, \tilde{{\cal L}}_{\phi}$, we obtain the stress tensor related to $\tilde{{\cal L}}_{\phi}$
\begin{equation}
\tilde{T}_{\mu \nu}=-\left[\phi_{,\mu}\phi_{,\nu}-g_{\mu \nu}\left(\frac{1}{2} g^{\alpha\beta}\phi_{,\alpha}\phi_{,\beta}-V(\phi)\right)\right],
\end{equation}
Since we are dealing with a spatially isotropic and homogeneous background metric (\ref{back}), the scalar field only depends on time, and comply with the dynamics
\begin{equation}\label{dy}
\ddot\phi + \left[3H(t)-\Gamma(t)\right]\dot\phi + \frac{\delta V(\phi)}{\delta\phi}=0,
\end{equation}
and the action written explicitly is
\begin{equation}\label{1}
{\cal I} = \int d^4x \, \sqrt{-\hat{g}} \,\left\{ \frac{{\cal R}}{16\pi G} - \left[\frac{\dot\phi^2}{2}\,e^{2\int \Gamma(t)\,dt} - V(\phi)\right]\right\},
\end{equation}
where the volume of the background manifold is $\hat{v}=\sqrt{-\hat{g}}=a^3_0\,e^{-\int \Gamma(t) \,dt}\,e^{3\int H(t)\, \,dt}$.
The action (\ref{1}) can be rewritten as\footnote{The new stress tensor is
${T}_{\mu \nu}=-\left[\phi_{,\mu}\phi_{,\nu}-g_{\mu \nu}\left(\frac{1}{2} g^{\alpha\beta}\phi_{,\alpha}\phi_{,\beta}-\bar{V}(\phi)\right)\right]$ and a new Lagrangian density corresponding to $\phi$: ${{\cal L}}_{\phi}= -\left[\frac{1}{2} g^{\alpha\beta} \phi_{,\alpha}\phi_{,\beta} - \bar{V}(\phi)\right]$, from which, due to the spatial isotropy and homogeneity, can be obtained explicitly the action (\ref{2}).}
\begin{equation}\label{2}
{\cal I} = \int d^4x \, \sqrt{-\hat{g}}\,e^{2\int \Gamma(t)\,dt}\,\left\{ \frac{\bar{\cal{R}}}{16\pi G} - \left[\frac{\dot\phi^2}{2} - \bar{V}(\phi)\right]\right\},
\end{equation}
where now the new scalar field $\phi$, will be solution of a new dynamics equation because it is embedded in an effective background volume $\hat{\bar{v}}=\sqrt{-\hat{g}}\,e^{2\int \Gamma(t)\,dt}=a^3_0\,e^{\int \Gamma(t)\, \,dt}\,e^{3\int H(t)\, \,dt}$. The redefined potential is $\bar{V}(\phi)=V(\phi)\,e^{-2\int \Gamma(t)\,dt}$, and there is an effective scalar curvature $\bar{\cal{R}}={\cal R}\,e^{-2\int \Gamma(t)\,dt}$. In this framework the stress tensor for the background inflation field, $\phi$, can be considered as a perfect fluid on an effective background volume $\hat{\bar{v}}$. We shall consider the model of an universe in which all the potential energy of the inflaton field is transferred to the expansion of this spacetime volume. In that case, the dynamics of the scalar field $\phi$ is given by
\begin{equation}
\ddot\phi + \left[3 H(t)+\Gamma(t)\right] \dot\phi + \frac{\delta \bar{V}}{\delta\phi} =0, \label{infl}
\end{equation}
that describes the dynamics of the background scalar field evolving on the background metric (\ref{back}). The first dissipative term is due to the expansion of the universe, but de second one is due to the existence of a nontrivial time scale. The background Einstein equations, are
\begin{eqnarray}
-3 H^2 \, e^{2\int \Gamma(t)\, \,dt}&=& -8\pi\, G \,T^0_{\,\,\,0}, \label{a} \\
-\left[3 H^2 + 2 \dot{H} + 2 \Gamma\, H \right]\,e^{2\int \Gamma(t)\, \,dt}\, \delta^{i}_{\,\,j} &=& -8 \pi \,G\,T^i_{\,\,\,j}, \label{b}
\end{eqnarray}
such that
\begin{equation}\label{stress}
P\,e^{2\int \Gamma(t)\, \,dt}\,\delta^{i}_{\,\,j}=-T^i_{\,\,\,j}=-\tilde{T}^i_{\,\,\,j}, \qquad \rho\,\, e^{2\int \Gamma(t)\, \,dt}=T^0_{\,\,\,0}=\tilde{T}^0_{\,\,\,0},
\end{equation}
where $P$ and $\rho$ are the solutions for $\Gamma=0$, for which we recover the expressions for a perfect fluid: $T^{\mu}_{\,\,\,\nu}=diag(\rho,-P,-P,-P)$. The results (\ref{stress}) can be obtained by using the relationship between $\bar{V}$ and $V$, and by equating the equations (\ref{infl}) and (\ref{dy}), so that the following conditions must be holds:
\begin{equation}
\Gamma(t) = \frac{\dot{V}}{2} \left[1 - e^{-2\int \Gamma(t)\, \,dt}\right].
\end{equation}
The last term in the left side of Eq. (\ref{b}) describes the contribution of the non-vacuum
to the dynamics of the field. The equation of state that describes the dynamics of the system is:
\begin{equation}
\omega= \frac{P}{\rho} = -1-\left(\frac{2 \dot{H}}{3H^2}+\frac{ 2 \Gamma }{3H}\right). \label{om}
\end{equation}
Notice that in the case of a vacuum expansion with $\Gamma=0$, the equation of state agrees with that of an ideal fluid. Of course, the most interesting case is $\Gamma\neq 0$. In that case it is possible, for example, to describe inflationary scenarios where dissipative effects are important in the evolution of the universe.

\subsection{Back-reaction effects}

A nonperturbative back-reaction formalism was developed in  earlier works\cite{rb1,rb2}. In those works it was demonstrated that the background metric can be altered by a scalar field $\sigma$ without the action to be altered: $\delta{\cal I}=0$, when it is fulfilled
\begin{equation}
-\frac{\delta \hat{\bar{v}}}{{\hat{\bar{v}}}}= \frac{ \delta \left[ \frac{\bar{{\cal R}}}{16\pi G} +{\cal L}_{\phi}\right]}{\left[ \frac{\bar{{\cal R}}}{16\pi G} +{\cal L}_{\phi}\right]} = 2\delta\sigma,
\end{equation}
where can be demonstrated that\cite{mb1}
\begin{equation}
\delta\sigma= -\frac{1}{2} g^{\alpha\beta} \delta g_{\alpha\beta}.
\end{equation}
Here, the back-reaction effects are due to the nonzero flux $g^{\alpha\beta} \delta R_{\alpha\beta}=-\frac{\Lambda}{2} \delta\sigma$, through a gaussian hypersurface, such that the manifold is defined by
\begin{equation}\label{ga}
\Gamma^{\alpha}_{\beta\gamma} = \left\{ \begin{array}{cc}  \alpha \, \\ \beta \, \gamma  \end{array} \right\}+\delta
\Gamma^{\alpha}_{\beta\gamma} =\left\{ \begin{array}{cc}  \alpha \, \\ \beta \, \gamma  \end{array} \right\}+ \,\sigma^{\alpha} g_{\beta\gamma} ,
\end{equation}
and the covariant derivative of the metric tensor, on this manifold, is
\begin{equation}\label{gab}
\delta g_{\alpha\beta} = g_{\alpha\beta|\gamma} \,dx^{\gamma} = {\nabla}_{\gamma}\, g_{\alpha\beta} \,dx^{\gamma}-\left[\sigma_{\beta} g_{\alpha\gamma} +\sigma_{\alpha} g_{\beta\gamma}
\right]\,dx^{\gamma},
\end{equation}
such that the covariant derivative of the metric tensor on the Riemannian manifold: ${\nabla}_{\gamma}\, g_{\alpha\beta}=0$, is zero. In other words, the metric tensor has null nometricity on the Riemann manifold, but not on the extended manifold defined by (\ref{ga}). From the point of view of the Riemann manifold $\Lambda$ is a constant, but from the point of view of the Weylian-like manifold: $\Lambda\equiv \Lambda(\sigma, \sigma_{\alpha})$ can be considered a functional, given by
\begin{equation}\label{aa}
\Lambda(\sigma, \sigma_{\alpha}) = -\frac{3}{4} \left[\hat{\Box} \sigma + \, \sigma_{\alpha} \sigma^{\alpha} \right].
\end{equation}
Therefore, a geometrical quantum action on the Weylian-like manifold with (\ref{aa}), can be considered
\begin{equation}
{\cal W} = \int d^4 x \, \sqrt{-\hat{g}} \,\,e^{2\int \Gamma(t)\,dt}\, \Lambda(\sigma, \sigma_{\alpha}),
\end{equation}
such that the dynamics of the geometrical field is given by the Euler-Lagrange equations, after imposing $\delta
W=0$. The dynamics of the back-reaction is described by the equation
\begin{equation}\label{si}
\ddot\sigma + \left[3H+\Gamma\right] \dot\sigma -\frac{1}{a^2_0} e^{-2\int \left[H+\Gamma\right]\,dt}\, \nabla^2 \sigma=0.
\end{equation}
Notice that the term $\Gamma \dot\sigma$ in (\ref{si}), takes into account the interaction between the geometric field $\sigma$ with the background. In order to describe the algebra of $\sigma$, we define the scalar invariant
$\Pi^2=\Pi_{\alpha}\Pi^{\alpha}$. If we require that $[\Pi^2,\sigma]=0$, we obtain the algebra\cite{rb1,rb2}
\begin{equation}\label{con}
\left[\sigma(x),\sigma^{\alpha}(y) \right] =- i \hbar\, \hat{U}^{\alpha}\, \delta^{(4)} (x-y), \qquad \left[\sigma(x),\sigma_{\alpha}(y) \right] =
i \hbar\, \hat{U}_{\alpha}\, \delta^{(4)} (x-y),
\end{equation}
where $\hat{U}^{\alpha}\equiv {d{x}^{\alpha}\over dS}$ are the components of the Riemannian
velocities, such that $g_{\mu\nu}\, U^{\mu}\,U^{\nu}=1$.

\section{An example: power-law inflation with variable timescale}

We consider the case where the Hubble parameter and the dissipative coefficient are
\begin{equation}
\Gamma(t) = p/t; \qquad H(t) = q/t,
\end{equation}
with $p$ and $q$ to be determined by observation parameters.
In this case the dynamics equation for the inflaton field holds
\begin{equation}\label{inf}
\ddot{\phi} + \left[\frac{3 q+p}{t}\right] \dot{\phi} + \frac{\delta \bar{V}}{\delta\phi} =0,
\end{equation}
where, in the Einstein equations (\ref{a},\ref{b}), we must set
$\,T^0_{\,\,\,0}=\left[\frac{\dot{\phi}^2}{2} + \bar{V}(\phi)\right]\, e^{2\int \Gamma(t)\, \,dt}$ and $\,T^i_{\,\,\,j}=-\delta^i_{\,\,j}\,\left[\frac{\dot{\phi}^2}{2} - \bar{V}(\phi)\right]\, e^{2\int \Gamma(t)\, \,dt}$. Notice that we are not considering radiation energy density. This is because all the energy of the inflaton field is transferred to expansion of the spacetime.

The geometrical scalar field $\sigma$ can be expressed as a Fourier expansion
\begin{equation}\label{four}
\sigma\left(\vec{x},t\right) = \frac{1}{(2\pi)^{3/2}} \int \, d^3k \, \left[ A_k \, e^{i \vec{k}.\vec{x}} \xi_k(t) + A^{\dagger}_k \, e^{-i \vec{k}.\vec{x}} \xi^*_k(t) \right],
\end{equation}
where $A^{\dagger}_k$ and $A_k$ are the creation and annihilation operators:
\begin{equation}\label{m5}
\left<B\left|\left[A_{k},A_{k'}^{\dagger}\right]\right|B\right>=\delta ^{(3)}(\vec{k}-\vec{k'}),\quad
\left<B\left|\left[A_{k},A_{k'}\right]\right|B\right>=\left<B\left|\left[A_{k}^{\dagger},A_{k'}^{\dagger}\right]\right|B\right>=0.
\end{equation}
The metric with back-reaction effects included, is
\begin{equation}\label{met1}
g_{\mu\nu} = {\rm diag}\left[e^{-2\int \Gamma(t)dt}\, e^{2\sigma}, - a_0^2 \,\, e^{2\int H(t) dt}\,\, e^{-2\sigma}, - a_0^2 \,\, e^{2\int H(t) dt}\,\, e^{-2\sigma}, - a_0^2 \,\, e^{2\int H(t) dt}\,\, e^{-2\sigma}\right],
\end{equation}
where the background scale factor $a(t)$ is given by (\ref{aa}). The relativistic quantum algebra is given by the expressions (\ref{con}), with co-moving relativistic velocities $U^0=e^{\int \Gamma(t)dt}$, $U^i=0$, which are calculated on the Riemannian (background) manifold.

Furthermore, as was calculated in a previous work\cite{Mio}, the variation of the energy density fluctuations is
\begin{equation}\label{de}
\left<B\left|\frac{1}{\bar{\rho}} \frac{\delta \bar{\rho}}{\delta S}\right|B\right> = - 2 \frac{\delta\sigma}{\delta S}= -2 \,U^0\, \sigma_0 =  -2 \,U^0\,\dot{\sigma},
\end{equation}
where $U^0= \left(\frac{t}{t_0}\right)^{p}$ and $\dot{\sigma} \equiv \left<B\left|\dot\sigma^2\right|B\right>^{1/2}$.

The equation of motion for the modes $\xi_k(t)$ in the expansion (\ref{four}), is
\begin{equation}\label{xi}
\ddot\xi_k(t) + \left[\frac{3 q+p}{t}\right] \dot\xi_k(t) + \frac{k^2}{a_0^2}\,\left(\frac{t}{t_0}\right)^{-2(p+q)} \xi_k(t)=0.
\end{equation}
Using the commutation relation (\ref{m5}) and the Fourier
expansions (\ref{four}) in
\begin{equation}
\left<B\left|\left[\sigma(t,\vec{x}), \Pi_{\alpha}(t,\vec{x}')\right]\right|B\right> = i\,\delta^{(3)}(\vec{x}-\vec{x}'),
\end{equation}
with canonical momentum: $\Pi_{\alpha}=\frac{\delta {\cal L}_{q}}{\delta \sigma^{\alpha}}=-{3\over 4} \sqrt{-\hat{g}}\,e^{2\int \Gamma(t) dt} \,\sigma_{\alpha}$,
we obtain the normalization condition for
the modes $\xi_{k}(\tau)$
\begin{equation}\label{m6}
\xi_{k}(t) \dot{{\xi}}^*_{k}(t) - \xi^*_{k}(t) \dot{{\xi}}_{k}(t) = i
\,\left(\frac{a^{3}_0}{e^{3\int H(t)\,dt}}\right),
\end{equation}
where the asterisk denotes the complex conjugated. The general solution of Eq. (\ref{xi}), is
\begin{equation}
\xi_k \left(t \right) ={t}^{-\frac{1}{2}\,\left(p+3\,q-1\right)} \,\left\{{\it A}\,
{\cal H}^{(1)}_{\nu}\left[y(t)\right]
+{\it B}\,{\cal H}^{(2)}_{\nu}\left[y(t)\right]\right\},
\end{equation}
where ${\cal H}^{(1,2)}_{\nu}\left[y(t)\right]$ are the first and second kind Hankel functions, with parameter
\begin{equation}
\nu={\frac {(p+3\,q-1)}{2(p+q-1)}},\label{nu}
\end{equation}
and argument $y(t)=\frac{k\,{{{t_0}}^{(p+q)}}{t}^{-(p+q-1)}}{{a_0}\, \left( p+q-1 \right) }$. To quantize we use the Bunch-Davies vacuum\cite{bd}, and we obtain
$\xi_k$, hold
\begin{equation}
\xi_k \left(t \right) =\sqrt{\frac{\pi}{4(p+q-1)}}\,\,{t}^{-\frac{1}{2}\,\left(p+3\,q-1\right)}\,{\cal H}^{(2)}_{\nu}\left[y(t)\right],
\end{equation}
and the power-spectrum on cosmological scales is
\begin{eqnarray}
{\cal P}_{\left<B\left|\frac{1}{\bar{\rho}} \frac{\delta \bar{\rho}}{\delta S}\right|B\right>}(k,t) & = &  \frac{1}{2\pi^2}  \frac{k^{n_s-1}}{\pi (p+q-1) (\beta t)^2}  \, \left(\frac{t}{t_0}\right)^{2p}\, \nonumber \\
& \times & \left[ \Gamma(\nu_1) \left[2(p-1) \beta\right]^{\nu_1} + [1-(3q+p)] \beta \Gamma(\nu) \, \left[2(p+q-1) \beta\right]^{\nu} \right]^2, \label{po}
\end{eqnarray}
with $\nu_1=\frac{3p+5q-3}{2(p+q-1)}$, and $\beta={a_0\, t^{-(p+q)}_0}$. The power of the spectrum is $3-2\nu$, which is related to the spectral index by the expression
\begin{equation}
3-2\nu = n_s-1.
\end{equation}
Furthermore, using the expression (\ref{om}), we obtain
\begin{equation}
\omega= \frac{-(2\nu+3)}{3(2\nu-1)},
\end{equation}
where $\nu$ is given by Eq. (\ref{nu}). Notice that for $p=1$, we obtain that $\nu=3/2$, $n_s=1$ and $\omega=-1$. This means that the model predicts a Harrison-Zel'dovich spectrum\cite{hz} for a vacuum expansion of the universe where the amplitude of the spectrum is frozen. However, last experimental data excludes $n_s=1$\cite{planck}, so that we shall contrast our model with the experimental data.

Now we consider the background dynamics given by the Einstein equations (\ref{a},\ref{b}), with the inflaton dynamics (\ref{inf}). From the Einstein equations, we can obtain
\begin{eqnarray}
\dot{\phi}^2 &=& \frac{1}{4\pi G} \left(\dot{H} + \Gamma H\right), \label{aa} \\
\bar{V} & = & \frac{1}{8\pi G} \left[3 H^2 + \left(\dot{H} + \Gamma H\right)\right]. \label{bb}
\end{eqnarray}
From (\ref{aa}), we obtain that
\begin{equation}
\dot{\phi}(t) = - \sqrt{\frac{q(1-p)}{4\pi\, G}} \,t^{-1}.
\end{equation}
Notice that the case $p=1$, corresponds to $\omega=-1$ and $\nu=3/2$, so that $n_s=1$, and the spectrum is scale invariant. In this case the power of the spectrum for back-reaction effects (\ref{po}) are independent of time, so that the amplitude of the back-reaction spectrum on cosmological scales is frozen. Furthermore, in this case $\dot{\phi}=0$, so that the inflaton field assumes a constant value.
On the other hand, due to the fact that $\frac{\delta V}{\delta\phi} = \dot{V}/{\dot{\phi}}$, using the expressions $H(t)=q/t$ and $\Gamma(t)=p/t$ in Eq. (\ref{inf}), we obtain the condition
\begin{equation}
q \left(p^2+3 p q + 3 q\right) =0,
\end{equation}
which gives us two possible solutions: $p=1-3q$ and $p=-1$. The case $p=1-3q$ is consistent with a decelerated expansion of the universe for $\nu=1.5175\pm 0.002$. The case $p=-1$, is consistent with a very accelerated expansion of the universe: $q=116$ and $\omega=-0.9885\pm 0.0013$, which is in very good agreement with observations\cite{planck}.
The scalar spectral index is $n_s=1-6\epsilon + 2 \eta$ and the tensor index is given by $n_t=-2\epsilon$, where the slow-roll parameters are given by the expressions\cite{slow}. In the second column of table \ref{tab1} we calculate the physical parameters for $n_s=0.965$ for $p=-1$:
\begin{equation}
\epsilon = 3\,\frac{\dot\phi^2}{2\bar{V}+\dot\phi^2}, \qquad \eta=-3 \,\frac{\ddot{\phi}}{(3H+\Gamma)\dot{\phi}} ,
\end{equation}
meanwhile the tensor-scalar ratio is given by $r=-8 \,n_t$. The observational cuts of these parameters from Planck 2018 results\cite{planck} are shown in the second column of the table (\ref{tab1}).
\begin{table}[htbp]
\begin{center}
\begin{tabular}{|l|l|l|}
\hline
Parameters &  For $n_s=0.968$  \\ \hline
\hline
$\epsilon=\left.{1-p\over q}\right|_{(p=-1)}={1-n_s\over (3-n_s)} $   & $ 0.01526 \, <\epsilon < \,0.01912 $  \\ \hline  \hline
$\eta=\left.{3\over 3q+p}\right|_{(p=-1)}={3(n_s-1)\over 5n_s-17} $  & $ 0.0076 \leq\eta \leq 0.0096 $  \\ \hline
$n_s=\left.{3p+q-3\over p+q-1}\right|_{(p=-1)}={q-6\over q-2}$  &  $ 0.961 \, < n_s < \, 0.969$  \\ \hline
$n_t=\left.-2\epsilon\right|_{(p=-1)}={-2(1-n_s)\over (3-n_s)}$  &  $ -0.0382 \,< n_t < \,-0.0305$  \\ \hline
$\left.r\leq 16 \epsilon\right|_{(p=-1)} ={16(1-n_s)\over (3-n_s)}$  & $ 0.2442\, <r <\,0.3060 $   \\ \hline
$q={2(3-n_s)\over 1-n_s} $  & $104.56 \, < q < \, 131.03$   \\ \hline
$\omega={n_s-7\over 3(3-n_s)} $  & $- 0.9872 \, < \omega < \, -0.9898$  \\ \hline
\end{tabular}
\caption{Observational cuts for slow-roll parameters for $n_s=0.965\pm 0.004$ and $p=-1$.}
\label{tab1}
\end{center}
\end{table}
In our model, $n_s \geq 1$ means that $\omega\leq -1$, which is excluded and therefore $p=-1$.

\section{Final Comments}

\begin{figure}[h]
\noindent
\includegraphics[width=.6\textwidth]{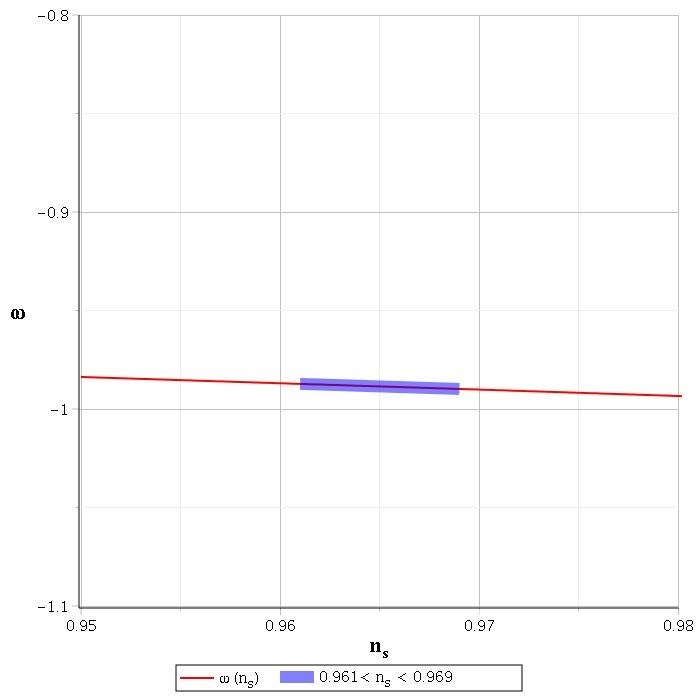}\vskip -0cm\caption{Plot of the range of $\omega(n_s)$-values for Planck 2018 data.}\label{f1gi108}
\end{figure}

\begin{figure}[h]
\noindent
\includegraphics[width=.6\textwidth]{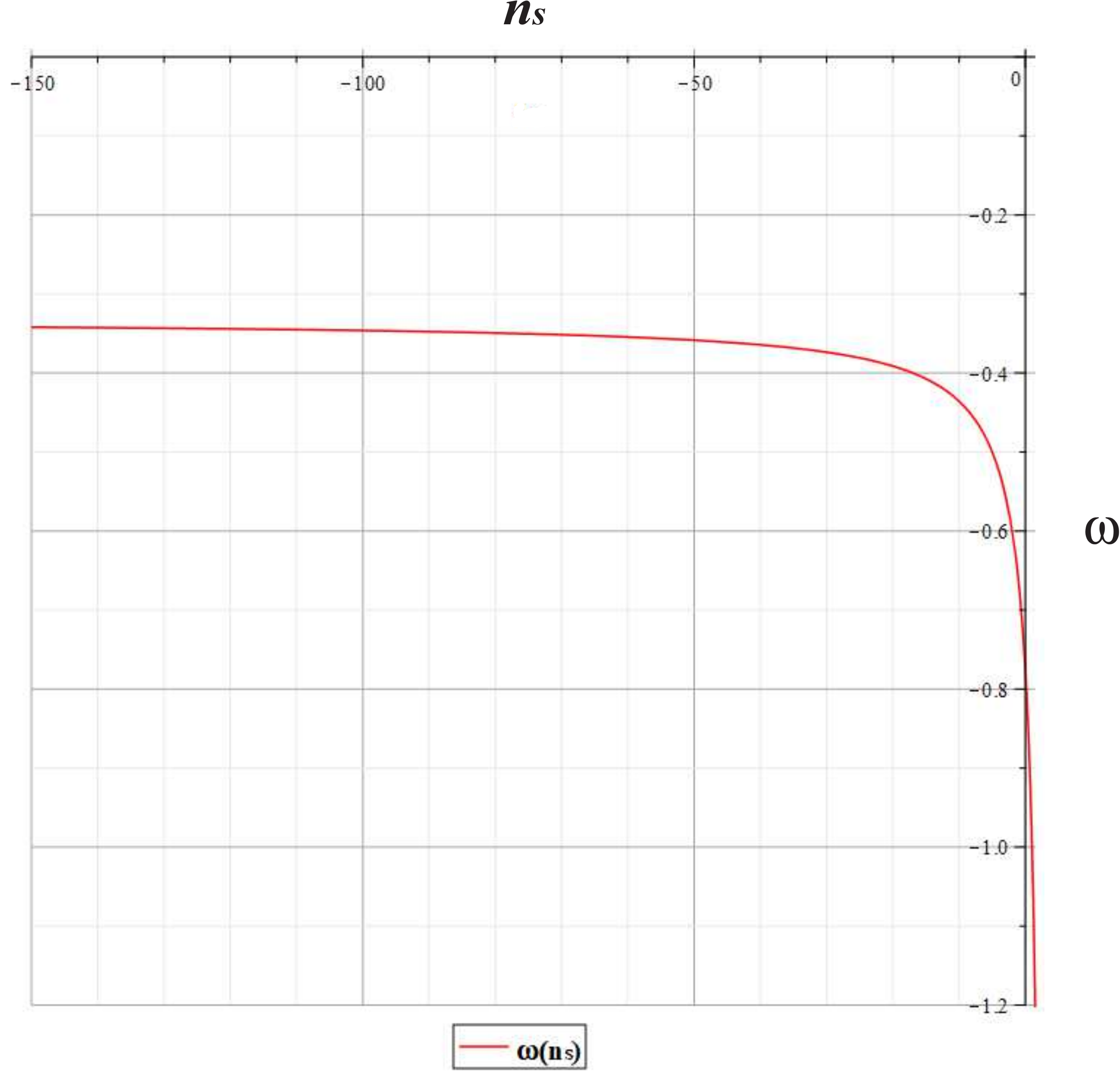}\vskip -0cm\caption{Plot of the range of $\omega(n_s)$-values for $\omega < -1/3$, in the range where the universe describes an accelerated expansion. }\label{f2gi108}
\end{figure}

We have studied a cosmological model in which the scale of time is variable and the expansion of the universe is driven by a scalar field. The dynamics of the scalar field together with the Einstein equations require that $p=-1$, so that the physical time is $\tau={1\over 2\, t_0} t^2 $. This means that at the beginning of the expansion the rate of events is much less hysterical, but after certain amount of expansion, the rate of co-moving events becomes more and more quickly, and we need much less physical time $\tau$ for a physical event to occur. From the point of view of a co-moving relativistic observer, its "clock" is accelerating with the expansion and the cosmic time, $t$, because the physical time evolves as $d\tau=U_0\,dx^0=\sqrt{g_{00}}dx^0=\left(t/t_0\right)\,dt$, for $c=1$. Beyond that, the proposed metric give us the possibility to describe exactly back-reaction effects for any equation of state. This is a great advantage with standard cosmological models where the timescale is not variable. In the Fig. (\ref{f1gi108}) we have plotted the plausible range of $q$ for $p=-1$, which corresponds to spectral indices $n_s=0.965\pm 0.004 $. The results show the range $ 0.961 \leq n_s \leq 0.969 $[see Fig. (\ref{f1gi108})], for which the rate of expansion of the universe is on the range, $104.56 < q < 131.03 $. In this range of $q$-values, $\omega$ takes the values: $- 0.9872 <\omega <-0.9898$. The results obtained for $n_s$, agree with a $k$-power of the spectrum (\ref{po}) that is close to zero, but negative. The amplitude for this spectrum decreases with time. In the Fig. (\ref{f2gi108}) we have plotted the curve of $\omega(n_s)$, for all possible values that describe an accelerated universe: $\omega < -1/3$. Our results exclude constraints from $WMAP9 + N_{eff}$ data\cite{ch} [with the effective number of relativistic species $N_{eff}$ and the massive neutrinos, simultaneously], but agree with constraints from Planck 2018 data\cite{planck}, which are compatible with $WMAP9$ data without consider the effective number of relativistic species and the mass of the neutrinos.

\section*{Acknowledgements}

\noindent The authors acknowledge CONICET, Argentina (PIP 11220150100072CO) and UNMdP (15/E810), for financial support.

\begin{thebibliography}{99}
\bibitem{var} F. Dowker, J. Henson and R. D. Sorkin, Mod. Phys. Lett. {\bf A19}: 1829 (2005); \\
L. Philpott, F. Dowker and R. D. Sorkin, Phys. Rev. {\bf D79}: 124047 (2009).
\bibitem{xz}  Yi-Fu Cai, Mingzhe Li, Xinmin Zhang, Phys. Lett. {\bf B718}: 248 (2012).
\bibitem{b} Mauricio Bellini, Phys. Dark Univ. {\bf 17}: 22 (2017).
\bibitem{lc} S. Liberati, Luca Maccione, Phys. Rev. Lett. {\bf 112}: 151301 (2014).
\bibitem{..} P. A. S\'anchez, M. Bellini, Int. J. Mod. Phys. {\bf D22}: 1342028 (2013).
\bibitem{da} G. R. Dvali, G. Gabadadze and M. A. Shifman, Phys. Lett. {\bf B497}: 271 (2001).
\bibitem{mb} M. R. A. Arcod\'{\i}a, L. S. Ridao, M. Bellini, Can. J. Phys. doi.org/10.1139/cjp-2018-0124.
\bibitem{mb1} M. R. A. Arcod\'{\i}a, M. Bellini, Universe {\bf 2}: 13 (2016).
\bibitem{rb1} L. S. Ridao, M. Bellini, Phys. Lett. {\bf B751}: 565 (2015).
\bibitem{rb2} L. S. Ridao, M. Bellini, Astrophys. Space Sci., {\bf 357}: 94 (2015).
\bibitem{Mio} M. Bellini, Phys. Dark Univ. {\bf 11}: 64 (2016).
\bibitem{bd} T. S. Bunch, P. Davies, {\it Quantum Field theory in de Sitter space: Renomarlization by point splitting}. Proc. Royal Soc. London {\bf A360}: 117 (1978).
\bibitem{slow} A. R. Liddle, P. Parsons, J. D. Barrow, Phys. Rev.{\bf D50}: 7222 (1994).
\bibitem{planck} N. Aghanim, {\it et all}, "Planck 2018 Results. VI. Cosmological parameters". E-print: arXiv: 1807.06209.
\bibitem{hz} E. R. Harrison, Phys. Rev. {\bf D1}: 2726 (1970); \\
Ya. B. Zel'dovich, NMRAS {\bf 160}: 1 (1972).
\bibitem{ch} Hong Li, Jun-Qing Xia, Xinmin Zhang, Phys. Dark Univ. {\bf 2}: 188 (2013).
\end{thebibliography}
\end{document}